\documentclass[twocolumn,english,eqsecnum,aps,pre,showpacs]{revtex4}
\usepackage[utf8]{inputenc}
\usepackage[english]{babel}

\usepackage{amsmath}
\usepackage{amssymb}

\makeatletter
\usepackage{graphicx}   % Include figure files
\usepackage{dcolumn}    % Align table columns on decimal point
\usepackage{bm}         % bold math
\usepackage{babel}

\begin{document}

\title{Non-equilibrium thermionic electron emission for metals at high temperatures
}

\author{J.L. ~Domenech-Garret }
\email{domenech.garret@upm.es}
\author{S.P. ~Tierno} 
\author{L.~Conde}
\affiliation{
 Departamento de F\'{\i}sica Aplicada,   
E.T.S.I. Aeron\'{a}utica y del Espacio.\\ 
Univ. Polit\'{e}cnica de Madrid, 28040 Madrid, Spain.}

\date{\today}

\begin{abstract}

The stationary thermionic electron emission currents from heated metals are compared against 
an analytical expression derived using a non equilibrium quantum Kappa energy distribution 
for the electrons. This later depends on the temperature decreasing parameter $\kappa(T)$ 
which can be estimated  from the raw experimental data and  characterizes the departure of 
the electron energy spectrum from the equilibrium Fermi-Dirac statistics. The calculations 
accurately predict the measured thermionic emission currents for both high and moderate 
temperature ranges. The Richardson-Dushman law governs the electron emission for large 
values of Kappa or equivalently, for moderate metal temperatures. The high energy tail  in the 
electron energy distribution function which develops at higher temperatures or lower Kappa 
parameters,  increases the emission currents well over the predictions of the classical expression.  
This analysis  also permits the quantitative estimation of the departure of the metal electrons from the equilibrium 
Fermi-Dirac statistics.     
\end{abstract}

\maketitle

\section{Introduction}
\label{intro}

The thermionic electron emission from metals at high temperatures is today on the basis of countless technical applications. Different materials are currently used as hot cathodes to produce beams of negatively charged particles in electron guns, plasma sources or microwave devices such as klystrons or traveling wave tubes \cite{Carr}, and vacuum thermionic energy conversion devices \cite{Smith}.

The stationary flow of thermionic electrons from the metal surface is currently 
calculated by the classical Richardson-Dushman (RD) equation. This model considers 
the thermal equilibrium between the electron gas and the metal lattice and makes 
use of the Fermi-Dirac distribution for the electron energy spectrum \cite{Lindsay}. The RD 
thermionic electron current density $J_{RD}(T)$ essentially relies on the work 
function $W_{f}$ of the metal and its temperature $T$, disregarding other important 
factors, such as the geometry \cite{Wysocki, Riffe} of the emitting surface or its physical 
state \cite{Paderno}.

However, the electron energy distribution in metals at high temperatures frequently 
differs from the Fermi-Dirac statistics. The stationary RD electron emission regime 
represents the final step after different energy thermalization processes with shorter 
time scales \cite{Riffe,Ferrini,Wendelen,Rethfeld}. This time dependent decay towards 
the stationary thermal equilibrium have been studied by exposing the metal surfaces to 
ultrashort laser pulses. The subsequent evolution of the energy spectrum is later 
monitored to determine the electron energy relaxation rates \cite{Ferrini,Rethfeld,Bezhanov}.

The average electron energy relaxes fast, within the femtosecond time scale by 
collisions between electrons. Therefore, the electron $T_{e}$ and metal $T$ 
temperatures differ within these short times scales. The thermalization between 
the electron gas and the metal lattice requires of few tens of picoseconds due 
to the larger mass of phonons \cite{Ferrini,Wendelen,Riffe,Rethfeld}. The RD 
regime is considered to provide an adequate description of the electron emission 
for long characteristic times when the electron gas and the metal lattice reach 
the thermal coupling $T = T_{e}$ \cite{Wendelen,Rethfeld,Corkum}. Nevertheless, 
this pure RD thermionic emission regime of electrons has not been observed because 
the energy of the laser pulse lies over damage thresholds of most metals \cite{Ferrini}. 

This physical description relies on the implicit assumption of the long time scale relaxation 
to a thermal equilibrium where the Fermi-Dirac statistics describes the electron energy 
spectrum \cite{Ferrini,Rethfeld}. This might be not always the case, even under an efficient 
energy transfer between the electron gas and metal lattice. For high temperatures the metal 
might remain out of equilibrium because additional energy exchange processes take place, 
such as the intense emission of electromagnetic radiation, the development strong surface 
thermal gradients or the nonuniform electron emission. The high energy interactions between 
the electrons and phonons couple the electron gas with the overheated metal lattice, producing 
groups of fast electrons \cite{Riffe,Rethfeld}. Therefore, even for long time scales, the electron 
energy spectrum might differ from the Fermi-Dirac energy distribution in hot metals.

Consequently, the equilibrium Fermi-Dirac statistics needs to be replaced to account for these 
high energy electron groups. The usual equilibrium statistical physics cannot cope with these 
non-conventional electron energy distribution functions  previously proposed in different fields 
\cite{Kamel}.  Namely,  the  derivation of a generalized Planck radiation law \cite{Ourabah1}, the 
description of high temperature Fermi gases \cite{Algin} or the out of equilibrium warm dense 
matter \cite{Clerouin}.

Recently, an analytical expression was derived for the thermionic current density  $J_{\kappa}(T)$, 
using a modified Kappa energy distribution for the electrons \cite{Domenech}. This formula corresponds 
to the high energy, non equilibrium Fermi-Dirac distribution \cite{Livadiotis,Treumann}. The index $\kappa$ 
of this distribution accounts for the high energy tail in the electron energy spectrum,  as well as the  
Kappa deformed statistics previously proposed  \cite{Kanidakis1}.

In this paper the measurements of the stationary thermionic emission currents from metals are 
found in good agreement with the theoretical predictions of Ref. \cite{Domenech}.  The RD  
expression underestimates the experimental emission currents for high metal temperatures, 
whereas  $J_{RD}(T)$ and $J_{\kappa}(T)$ agree for moderate values.

 As we shall see, the electron energy distribution might be approximated using the temperature 
 dependent $\kappa(T)$ index of this statistics, which could be determined from the experimental 
data. This $J_{\kappa}(T)$ accounts for the contribution of high energy electrons in the thermionic electron emission.

Additionally, it might be employed as a quantitative estimation of the the departure from the thermal 
equilibrium described by the Fermi-Dirac statistics which is recovered in the limit of large $\kappa$.  
To our best knowledge, this present analysis is one of the few cases where the predictions of a non equilibrium quantum 
Kappa distribution are directly compared against raw experimental data. 
%%% FIGURA 0 %%%%%
\begin{figure}
\includegraphics[width=8.0cm]{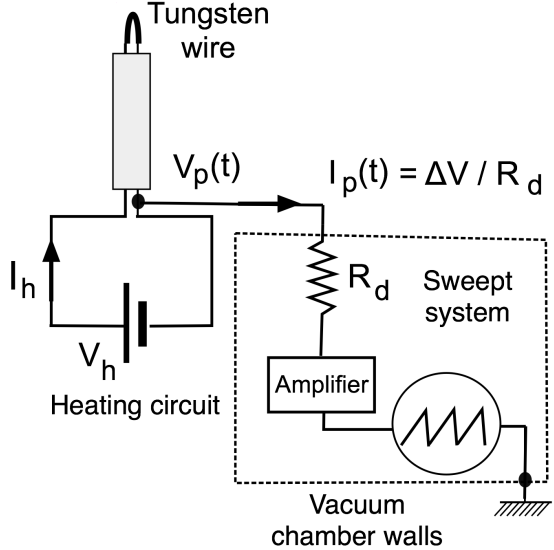}  
\caption{\label{fig:0} The diagram of experimental setup where,  $V_{h}$, is the voltage drop across the tungsten 
wire heated by the DC current, $I_{h}$. The dotted box represents the scheme of the sweep bias circuit 
which applies an amplified sawtooth waveform $V_{p}(t)$ to the heated wire and measure the emission 
current, $I_{p}(t)$,  as the voltage drop, $\Delta V$, across a precision resistor, $R_{d}$.}
\end{figure} 
%%%%%%%%%%%%%%%%%%%

\section{The non equilibrium electron emission model}

The stationary thermionic electron current from metals at high temperatures 
could be calculated using the stationary, quantum non-equilibrium Kappa statistics 
\cite{Domenech},  

\begin{equation}
f_{\kappa} (T_{e}, E) = C_{\kappa}(T_{e}) \, 
\left( 1 + \frac{E-\epsilon_{F}}{k_{B}T_{\kappa}}\right)^{-(\kappa+1)} 
\label{eq:KappaDist}
\end{equation}

\noindent 
Here $k_{B}$ is the Boltzmann constant and $E$ is the kinetic energy of electrons 
of mass $m_e$, which is evaluated with respect to the Fermi level $\epsilon_{F}$. 
The other parameters are \mbox{$\gamma_{1} = \epsilon_{F} / k_{B} T_{e}$} and  
\mbox{$T_{\kappa} = (\kappa - 3/2 + \gamma_{1} )\, T_{e}$}. Eq. (\ref{eq:KappaDist}) 
is the approximation for high energies of the generalized non equilibrium quantum 
statistics of Refs. \cite{Livadiotis} and \cite{Treumann}.
The normalization factor $C_{\kappa}(T_{e})$ in Eq. (\ref{eq:KappaDist}) scales to the metal 
electron density $n_{eo}$ as,

\begin{eqnarray}
C_{\kappa} (T_{e}) =\  n_{eo}\  
\frac{\Gamma(\kappa +1)}{\Gamma(\kappa - \frac{1}{2})}\times \nonumber \\
\times 
\biggl( \frac{m_e}{(\kappa-3/2)\ 2 \pi\ k_{B} \, T_{e}}\biggr)^{3/2}    
\biggl( \frac{\kappa-3/2}{\kappa-3/2 + \gamma_{1}} \biggr)^{\kappa+1}
\nonumber
\end{eqnarray} 

The low values of $\kappa$ in  Eq. (\ref{eq:KappaDist}) account for a large high 
energy tail in the electron energy spectrum, whereas large indexes $\kappa$ recover 
the classical Fermi-Dirac statistics \cite{Domenech}. 

%%% FIGURA 1 %%%%%
\begin{figure}
\includegraphics[width=8.0cm]{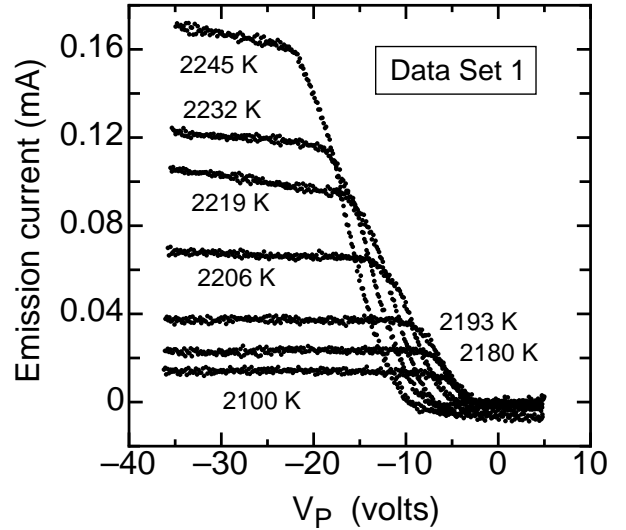}  
\caption{\label{fig:1} The electron emission currents $I_{e}(T)$ as a function 
of the bias potential $V_{p}$ for different metal temperatures.
}
\end{figure} 
%%%%%%%%%%%%%%%%%%%

For long time scales, well over the microsecond range equal temperatures \mbox{$T_{e} = T$} 
are considered for the electron gas and the metal lattice \cite{Ferrini,Rethfeld,Corkum}. The 
average electron thermal 
energy proportional to $k_{B}T_{e}$ is always much lower than the work function, $W_{f} $. 
In these conditions the analytical expression for the stationary thermionic electron current 
density $J_{\kappa}(T)$ deduced from Eq. (\ref{eq:KappaDist}) reads \cite{Domenech}, 

\begin{eqnarray}
J_{\kappa} (T) =\ e\  C_{\kappa} (T)\, \frac{\pi\ k^2_{B} \, T^2 }{m_e^2} \times \nonumber \\ 
\frac{\ (2\kappa-3+ 2 \, \gamma_{1})^2}{2 \kappa\ (\kappa-1)}  \  
\times \biggl( 1\ + \frac{W_{f}}{ \, k_{B} \,T_{\kappa}} \biggr)^{-\kappa+1}
\label{eq:Jk}
\end{eqnarray}

\noindent
For low $\kappa$ the thermionic electron fluxes predicted by  Eq. (\ref{eq:Jk}) 
are higher than those calculated using the classical Richardson-Dushmann expression 
$J_{RD}(T)$, which is also recovered in the opposite limit of large $\kappa$ 
\cite{Domenech}. 

The contribution in $J_{\kappa}(T)$ of the high energy tail of the electron energy distribution 
function increases with the metal temperature. This effect is incorporated by means of a 
temperature dependent index $\kappa(T)$ in either $f_{\kappa} (T, E)$ and $J_{\kappa}(T)$.

\section{Experiments}

The predictions of Eq. (\ref{eq:Jk}) were checked against the measurements of the thermionic electron currents $I_{e}(T,V_{p})$ from a DC heated tungsten and in Fig. \ref{fig:0} is depicted a simplified scheme of the experimental setup. The loop shaped wire with a typical length of 22 mm and 0.08 mm in diameter was placed at the end of a electrically insulated ceramic shaft.  The emitting wire was approximately located at the center of a cylindrical vacuum chamber of 0.8 m in length and 0.4 m in diameter evacuated down to 
typical pressures below \mbox{$ 10^{-5}$ mB} of Argon.  The elastic collisions between electrons and neutral  is the dominant collisional process. According to \cite{Raju} the cross sections are $9-20 \times 10^{-20}$ m$^{2}$, which gives typical mean free paths  between 20 and 44 m, much larger than our experimental  arrangement.  This fact excludes the production of additional charged particles because the ionizing  and elastic collisions between electron and neutrals are negligible.

The tungsten wire was heated up to thermionic electron emission by a DC currents of $I_{h} \simeq 0.8-1.0$ A with voltages $V_{h} \simeq 2-3 V$ and electrically biased with respect to the grounded walls of the 
vacuum tank. For these thin tungsten filaments the temperature gradients along the hot wire could be neglected and energy power losses are mainly caused by radiation emission. Therefore, the temperatures were determined using a well known expression relating the DC heating current along the wire with $T$, which is regarded essentially uniform \cite{Langmuir16,Halas,IE3Tierno}.

The sweep system  impress an amplified time dependent sawtooth signal, $V_{p}(t)$, with a repetition pulse of 2 kHz to one leg of the heated tungsten wire \cite{Troll}. As shows Fig. \ref{fig:0} the emitted thermionic electron current $I_{e}(T,V_{p})$ is obtained from the current, $I_{p}(t)$, measured as the voltage drop, $\Delta V$, across the precision resistance, $R_{d}$.

The typical electron emission currents $I_{e}( T, V_{p})$ are 
represented in Fig. \ref{fig:1} for different metal temperatures $T$ as a function 
of the bias potential $V_{P}$ of the wire. For negative bias potentials $V_{p}$ the 
thermionic electrons are emitted from the wire towards the vacuum chamber walls, 
whereas no electron current is observed for positive potentials.

%%% FIGURA 2 %%%%%
\begin{figure}
\includegraphics[width=8.0cm]{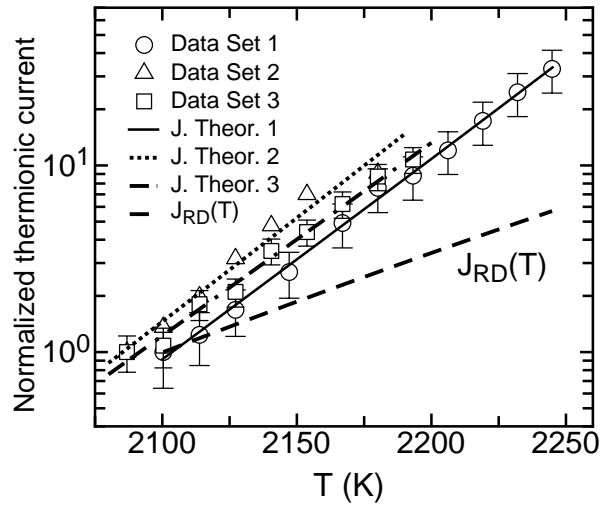}  
\caption{\label{fig:2} Comparison between the experimental data $I_{s}(T)/I_{s}(T_{o})$ and  the 
normalized thermionic current densities $j_{\kappa}(T) = J_{\kappa}(T)/J_{\kappa_{m}}(T_{o})$ 
(labeled as J. Theor.) by Eq. (\ref{eq:Jk}) using the fitting $\kappa(T)$ of Fig. \ref{fig:3}. The dashed line represents the predictions of the normalized Richardson-Dushmann equation $j_{RD}(T)$ as a function of the metal temperature.} 
\end{figure} 
%%%%%%%%%%%%%%%%%%%

The lower currents in Fig. \ref{fig:1} were of tens of $\mu$A corresponding 
to \mbox{$T <$ 2100 K} that raised up to the \mbox{0.1-0.2 mA} range for metal 
temperatures over \mbox{2200 K}. For low positive potentials where the electron 
emission becomes negligible, $I_{e}( T, V_{p})$ reaches a flat negative current for voltages over a threshold of about 5 volts. The thermionic electron emission takes place for negative bias potentials and large slopes in the range -5 to -20 volts correspond to the space charge effects around the wire.  These are outweighed over a typical threshold bias potential of the wire corresponding to the knee of $I_{e}(T, V_{p})$ where the flat electron emission current becomes weakly dependent on the bias potential. 

Then, the flat saturation currents \mbox{$I_{s}(T) \simeq I_{e}( T, V_{p})$}
for \mbox{$V_{p} \leq -20$} volts past the knee of $I_{e}( T, V_{p})$, measure the maximum 
emitted electron thermionic current. The moderate bias voltages $V_{p}$ involved 
in Fig. \ref{fig:1} exclude from this pure thermionic regime the thermofield or 
Fowler-Nordheim or field emission modes \cite{Murphy,Coulombe}.

\section{Experimental data analysis}

The theoretical predictions of  Eq. (\ref{eq:Jk}) are compared with the  measurements of $I_{s}(T)$ in  Fig. \ref{fig:2}. For low metal temperatures (typically below 2000 K) both, the RD expression and Eq. (\ref{eq:Jk}) give similar thermionic electron emission currents. 
However, the increment in the metal temperature leads to important differences. 

\begin{figure}
\includegraphics[width=8.0cm]{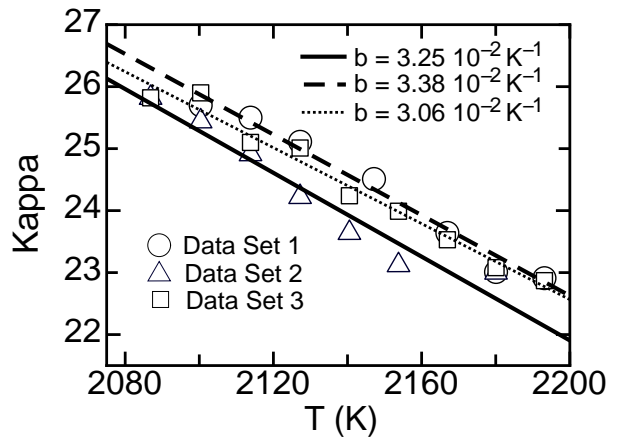}  
\caption{\label{fig:3} The parameter $\kappa(T)$ corresponding to the experimental data fitting, using the least-squares method, of the three data sets of Fig. (\ref{fig:2}) to $\kappa(T) = a - b \, T$, 
where $T$ is the metal temperature. 
}
\end{figure}

Additionally, the values of index $\kappa(T)$ of $f_{\kappa} (T, E)$ could be evaluated 
from the maximum thermionic electron emission currents $I_{s}(T)$  and are 
represented in Fig. \ref{fig:3}.  The low temperature limit corresponds in Fig.  \ref{fig:2} 
to large values of the index $\kappa(T)$ where  $f_{\kappa} (T, E)$ becomes similar to the 
equilibrium Fermi-Dirac statistics. The decreasing $\kappa(T)$ index characterizes the 
 $f_{\kappa} (T, E)$ for growing metal temperatures.
 
The following procedure was used in Figs. \ref{fig:2} and \ref{fig:3} to  avoid the need of 
the accurate determination of the electron emitting metal surface, as well as the particular 
value of the Richardson constant for the metallic samples.

First, we obtain the parameter $\kappa_{m}$ for the lowest metal temperatures 
\mbox{$T_{o} \simeq $ 2100 K} in Fig. \ref{fig:1}, which corresponds to the Richardson-Dushman electron emission regime. From  Eq. (\ref{eq:Jk}) and 
the classical RD expression we obtain  $k_{m}$ as the root of a nonlinear equation by setting, 
$$
\frac{J_{\kappa_{m}}(T_{o})}{J_{RD}( T_{o})} = 1 
$$ 
\noindent
This permits to evaluate the lower current density $J_{\kappa_{m}}(T_{o})$ using Eq. (\ref{eq:Jk}) which corresponds to the lower measured thermionic current $I_{s}(T_{o})$ of Fig. \ref{fig:1}. The 
maximum values is about  \mbox{$k_{m} \simeq 25.6-26.0$} in  Fig. \ref{fig:3} for the three sets of experimental data. 

Next, using the maximum thermionic emission currents $I_{s}(T)$ for the metal temperature $T$ 
the ratios $I_{s}(T)/I_{s}(T_{o})$ are therefore,
$$
  \frac{I_{s}(T)}{I_{s}(T_{o})} = \frac{J_{\kappa}(T)}{J_{\kappa_{m}}(T_{o})} = j_{\kappa}(T)
$$

\noindent
Using Eq. (\ref{eq:Jk})  the values $\kappa(T) <  \kappa_{m} $ for increasing temperatures 
$T > T_{o}$ could be again  evaluated as the roots of a nonlinear equation. 

Finally, the temperature decreasing values of  $\kappa(T)$ of Fig. \ref{fig:3} could be approximated by  $\kappa(T) = a - b\,T$ by means of a least-squares fitting and these  empirical expressions are later introduced in Eq. (\ref{eq:Jk}). This procedure allows to compare its  theoretical predictions with the experimental ratios $I_{s}(T) / I_{s}(T_{o})$. 
The normalized curves  $j_{\kappa}(T)$ along with the experimental ratios  $I_{s}(T)/I_{s}(T_{o})$ are 
represented in Fig. \ref{fig:3} as well as,
$$
j_{RD}(T) = \frac{J_{RD}(T)}{J_{RD}(T_{o})}
$$
\noindent
which it were calculated by using the classical Richardson-Dushmann equation.

%%% FIGURA 3 %%%%%
%%%%%%%%%%%%%%%%%%%

\section{Conclusions}

The classical Richardson-Dushman equation based on the equilibrium Fermi-Dirac statistics
adequately describes the thermionic electron emission for low and moderate metal temperatures. 
However,  additional energy exchange processes bring the metal lattice far form the thermal 
equilibrium at higher temperatures. Thus, a fraction the lattice energy is also transferred to the 
electron gas for long time scales. In these conditions, the electron energy spectrum differs from 
the Fermi-Dirac statistics and could be approximated by the effective, non equilibrium quantum 
Kappa distribution of Eq. (\ref{eq:KappaDist}). This later recovers the Fermi-Dirac statistics for 
low temperatures or equivalently,  for large values of the $\kappa(T)$ index in $f_{\kappa}( T, E)$. 

The empirical determination of the index $\kappa(T)$ from the stationary thermionic electron 
emission currents $I_{s}(T)$ for different metal temperatures was discussed in Sec. IV.  The 
maximum value $\kappa_{m}$  for the lower temperature in  Fig. \ref{fig:3} corresponds to 
the RD regime and the decreasing function $\kappa(T)$ quantitatively characterizes the departure 
of the electron energy spectrum  from  the Fermi-Dirac statistics \cite{Domenech}.  This transition 
rate is governed by the slope $b$ in the empirical fit to  to $\kappa(T) = a - b \, T$ of Fig. \ref{fig:3} 
which  also  characterize the increasing high energy tail  in  $f_{\kappa} ( T, E)$  with the metal  
temperature.

Fig. \ref{fig:2} shows the good agreement between the experimental data and the thermionic 
electron emissions currents $J_{\kappa}(T)$ calculated using $f_{\kappa}( T, E)$ for the electron 
energy  spectrum and the empirical expressions for $\kappa(T)$. Additionally, the measurements 
of Fig. \ref{fig:2} show the increase of the thermionic current by orders of magnitude over RD law 
predictions for high metal temperatures. These increments are  caused by the departure for the 
thermal equilibrium of both the metal lattice and the electron gas not contemplated by the equilibrium 
Fermi-Dirac statistics employed in the derivation of the classical RD equation. 

This high thermionic electron emission regime is  explained by the contribution of high energy 
electron groups in the energy spectrum of the metal electrons. Therefore, 
Eq. (\ref{eq:Jk}) extends the predictions of the classical Richardson-Dushman equation to higher metal temperature 
and permits to make us of $\kappa(T)$  to approximate the stationary energy distribution function 
of electrons in hot metals. 

Finally, regarding the applicability of these results, cathodes with higher thermionic electron emission currents would led to improved characteristics in electron tube devices. Larger electron currents lead to greater amplification factors in traveling wave tubes or a better coupling with the microwave input signal in klystrons. 
   
\begin{acknowledgments}
The authors are grateful to the MICINN funding through Grant ESP2013-41078-R and 
S.P. Tierno acknowledges the financial support through FPU program from the Spanish 
Ministry of Education.
\end{acknowledgments}

%%%%%%%%%%%%%%%%%%%%%%% 60 CARACTERES %%%%%%%%%%%%%%%%%%%%%


\begin{thebibliography}{1}


\bibitem{Carr} J.J.~Carr, 
             \textit{Microwave \& Wireless Communications Technology}
              (Elsevier, New York, 1996), Chap. 11.
              
\bibitem{Smith}  J.R.~Smith,
                 J. Appl. Phys. \textbf{114}, 164514 (2013).


\bibitem{Lindsay} R.B.~Lindsay, 
              \textit{Introduction to Physical Statistics} 
              (Wiley, New York, 1962) Chap. 11. 
              
\bibitem{Wysocki} J.K.~Wysocki, Phys. Rev. B \textbf{28}, 834 (1983).

\bibitem{Riffe}  D.M. Riff, X.Y. Wang, M.C. Downer, D.L. Fisher, T. Tajima,  J.L. Erskine, and R.M. More.         
             J. Opt. Soc. Am. B. \textbf{10}, 1424 (1993).                                                      

\bibitem{Paderno} Y.U.~Paderno, A.A.~Taran, D.A.~Voronovich, V.N.~Paderno and V.B.~Filipov,    
                     Functional Materials {\bf 15}, 63  (2008).   
                     
\bibitem{Ferrini} G.~Ferrini, F.~Banfi, C.~Giannetti and F.~Parmigian,
                   Nuc. Inst.  Meth. Phys. Res. A,  \textbf{601}, 123 (2009). 

\bibitem{Wendelen} W.~Wendelen, B.Y.~Mueller, D.~Autrique, B.~Rethfeld and A.~Bogaerts,
                   J. Appl. Phys. \textbf{111}, 113110 (2012).

\bibitem{Rethfeld}  B.~Rethfeld, A.~Kaiser, M.~Vicanek and G.~Simon.
                 Phys. Rev. B, \textbf{65}, 214303 (2002).

\bibitem{Bezhanov} S.G. Bezhanov, A.P. Kanavin and S.A. Uryupin,
              Quantum Electronics \textbf{42},  447 (2012).
              
\bibitem{Corkum} P.B. Corkum, F. Brunel and  N.K. Sherman, T. Srinivasan-Rao,
             Phys. Rev. Lett. \textbf{61},  2886 (1988).
             
\bibitem{Kamel} K.~Ourabah, L.~Ait Gougam and M.~Tribeche,
                Phys. Rev. E, \textbf{91}, 012133 (2015).

\bibitem{Ourabah1} K.~Ourabah and M.~Tribeche,
                Phys. Rev. E, \textbf{89}, 062130 (2014).

 \bibitem{Algin} A. Algin and M. Senay, 
                Phys. Rev. E \textbf{85}, 041123 (2012).
                
\bibitem{Clerouin} J.~Cl\'erouin, G.~Robert, P.~Arnault, C.~Ticknor, J.D.~Kress and L.A.~Collins,
                 Phys. Rev. E, \textbf{91}, 011101(R) (2015).
                              
\bibitem{Domenech} J.L.~Domenech-Garret, S.P.~Tierno and L.~Conde, 
              Eur. Phys. J. B, {\bf 86}, 382 (2013).                 
                 
 \bibitem{Livadiotis} G.~Livadiotis, D.J.~McComas, 
               Astrophs. J. 741:88 (2011).
                                         
\bibitem{Treumann}   R.A.~Treumann,   Europhys. Lett., \textbf{48},  8 (1999).  
                
\bibitem{Kanidakis1} G.~Kaniadakis, Physica A {\bf 296}, 405 (2001).


\bibitem{Raju} G.G.~Raju,  
        IEEE Trans. on Dielectrics and Electrical Insulation {\bf 11}(4),
        649 (2004), Figure 1 Page 654.  

\bibitem{Langmuir16} I.~Langmuir, Phys. Rev. 7, 302 (1916).

\bibitem{Halas} S.~Halas and T.~Durakiewicz, 
Vacuum, \textbf{49}, (4) pp. 331-336 (1998).

\bibitem{IE3Tierno} S.P.~Tierno, J.L.~Domenech-Garret, J.M.~Donoso, D.~Jennewein, G.~Herdrich, S.~Fasoulas, and  L.~Conde, 
IEEE Trans. Plasma Sci. {\bf 41}, 695 (2013).

\bibitem{Troll} O.~Troll, L.~Conde, E.~Criado, J.M.~Donoso, and G.~Herdrich. 
                 Contrib. to Plasma Phys. {\bf 50}(9), 819 (2010).

               
\bibitem{Murphy} E.L. Murphy and R.H. Good, 
              Phys. Rev. \textbf{102}, (6) 1464 (1956).

\bibitem{Coulombe} S. Coulombe and J.L. Meunier, 
             J. Phys. D: Appl. Phys. \textbf{30} 776 (1997).
           							

\end{thebibliography}
\end{document}